\documentclass[pra,aps,superscriptaddress,nofootinbib,twocolumn]{revtex4-1}

\usepackage{mathrsfs}
\usepackage{amsfonts}
\usepackage{amssymb}
\usepackage{amsmath}
\usepackage{graphicx}
\usepackage[usenames,dvipsnames]{color}
\usepackage[colorlinks=true,citecolor=blue,linkcolor=magenta]{hyperref}
\usepackage{ulem}
\usepackage{lmodern}
\usepackage{amsthm}

\newcommand{\figpath}{.}

\newcommand{\Tr}{\mathrm{Tr}}

\newcommand{\abs}[1]{\vert #1 \vert}

\newcommand{\ket}[1]{\vert{ #1 }\rangle}

\newcommand{\ketbra}[2]{\vert #1 \rangle \langle #2 \vert}
\newcommand{\braket}[2]{\langle #1 \vert #2 \rangle}
\newcommand{\mean}[1]{\langle #1 \rangle}

\begin{document}

\title{Simulation of memristive synapses and neuromorphic computing on a quantum computer}

\author{Ying Li}
\affiliation{Graduate School of China Academy of Engineering Physics, Beijing 100193, China}

\begin{abstract}
One of the major approaches to neuromorphic computing is using memristors as analogue synapses. We propose unitary quantum gates that exhibit memristive behaviours, including Ohm's law, pinched hysteresis loop and synaptic plasticity. Hysteresis depending on the quantum phase and long-term plasticity that encodes the quantum state are observed. We also propose a three-layer neural network with the capability of universal quantum computing. Quantum state classification on the memristive neural network is demonstrated. Our results pave the way towards brain-inspired quantum computing. We obtain these results in numerical simulations and experiments on the superconducting quantum computer {\it ibmq{\textunderscore}vigo}. 
\end{abstract}

\maketitle

\section{Introduction}

\begin{figure}[tbp]
\centering
\includegraphics[width=1\linewidth]{\figpath 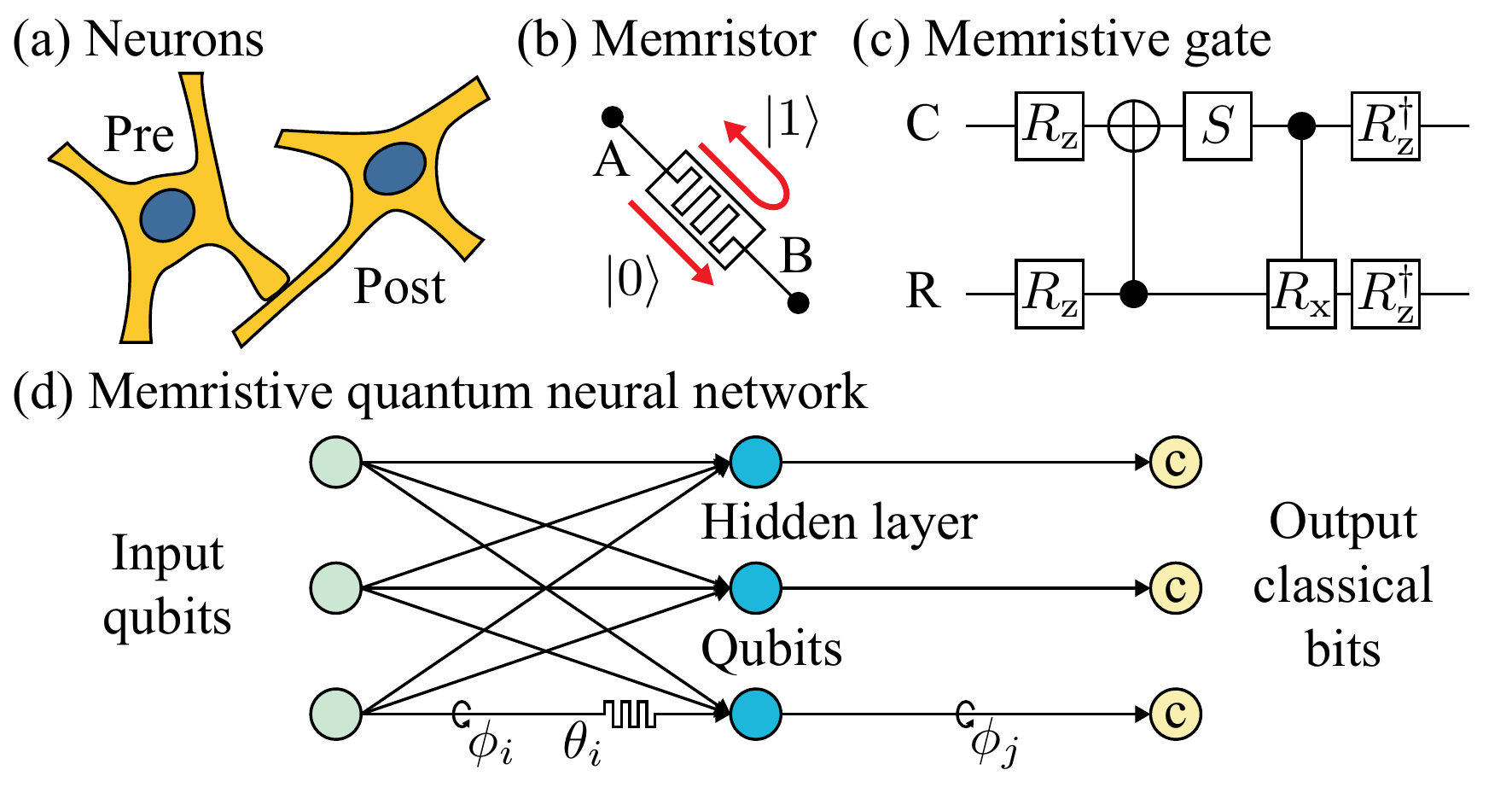}
\caption{
(a) Pre- and post-synaptic neurons. 
(b) A memristor. In the quantum regime, we use qubits to represent the input/output current and the resistance of the memristor. 
(c) Memristive gate $M_\theta$ decomposed into elementary quantum gates, where $R_{\rm z} = e^{-i\frac{\theta}{2}Z}$ and $R_{\rm x} = e^{-i(\frac{\pi}{2}-\theta)X}$. 
(d) Memristive quantum neural network. 
}
\label{fig:neural_network}
\end{figure}

Neuromorphic computing is a brain-inspired computer paradigm in contrast with the von Neumann architecture~\cite{Mead1990, Schuman2017}. According to the biological model of the brain, the information is stored and processed by a highly connected network formed of neurons, which provides the ability of learning, parallel and low energy cost computing, etc. Since the 1940s, it has been realised that how neurons wire up is essential~\cite{Hebb1949}. Besides neuroscience, this observation also motives the development of computer programming, such as the neural network algorithms vastly used in today's machine learning technologies~\cite{Nielsen2015, Goodfellow2016}. In term of the learning rule of neurons, spike-timing-dependent plasticity (STDP) is a biologically plausible model that has gained great attention in recent years~\cite{Caporale2008, Markram2011, Feldman2012}. In STDP, the synapse is strengthened or weakened depending on the temporal order between spikes of pre- and post-synaptic neurons [see Fig.~\ref{fig:neural_network}(a)]. In this way, the brain can establish causal relationships between events. 

Quantum computing uses quantum phenomena and is superior to classical computing in solving certain problems~\cite{Nielsen2010}. For example, to solve the integer factorisation problem, Shor's quantum algorithm takes polynomial time with respect to the integer size, which is exponentially faster than the most efficient known classical algorithm~\cite{Shor1994}. In the circuit-based universal quantum computer, information is encoded in qubits and processed with unitary gates~\cite{Deutsch1985}. This kind of quantum machines is still under development but already demonstrates the power of surpassing classical computers~\cite{Google, IBM}. Because the quantum computer for large-scale computing is not available yet, variational quantum algorithms are proposed for the near-future applications~\cite{Peruzzo2014, Farhi2014, Li2017}. Quantum neural networks are generalisations of classical artificial neural networks, in which unitary gates in the quantum circuit are taken as variables~\cite{Beer2020, Wan2017, Romero2017, Cao2017, Farhi2018, Mitarai2018, Grant2018, Schuld2020, Killoran2019, Steinbrecher2019}. 

The memristor is a resistor with memory and one of the fundamental two-terminal circuit elements~[see Fig.~\ref{fig:neural_network}(b)]~\cite{Chua1971, Chua1976}. Its resistance decreases or increases depending on the input signal, i.e.~the voltage or current. Memristance can explain STDP in biological synapses~\cite{LinaresBarranco2009}. Since the first memristive device was found in 2008~\cite{Strukov2008}, the application as hardware analogue of synapse in neuromorphic computing has been extensively investigated~\cite{Schuman2017}, mainly because memristive devices demonstrate behaviours similar to STDP~\cite{Jo2010, Serb2016}. 

In this paper, we propose memristor-like unitary quantum gates. These gates have the characteristic memristive property, i.e.~hysteretic resistance state~\cite{Chua1976, Strukov2008}. Given an oscillatory input state, the output-input observables display a pinched hysteresis loop. We find that the loop depends on not only the classical distribution but also the phase of the input quantum state, which reflects the quantum nature of memristive gates. Using these gates to mimic synapses, we observe the long-term potentiation (LTP) and long-term depression (LTD), which are crucial for learning and memory in the neural network~\cite{Caporale2008, Feldman2012}. We show that quantum information can also be encoded in a manner similar to the long-term plasticity. Therefore, a neuromorphic computer based on the memristive gates can process quantum information. 

\begin{figure*}[tbp]
\centering
\includegraphics[width=1\linewidth]{\figpath 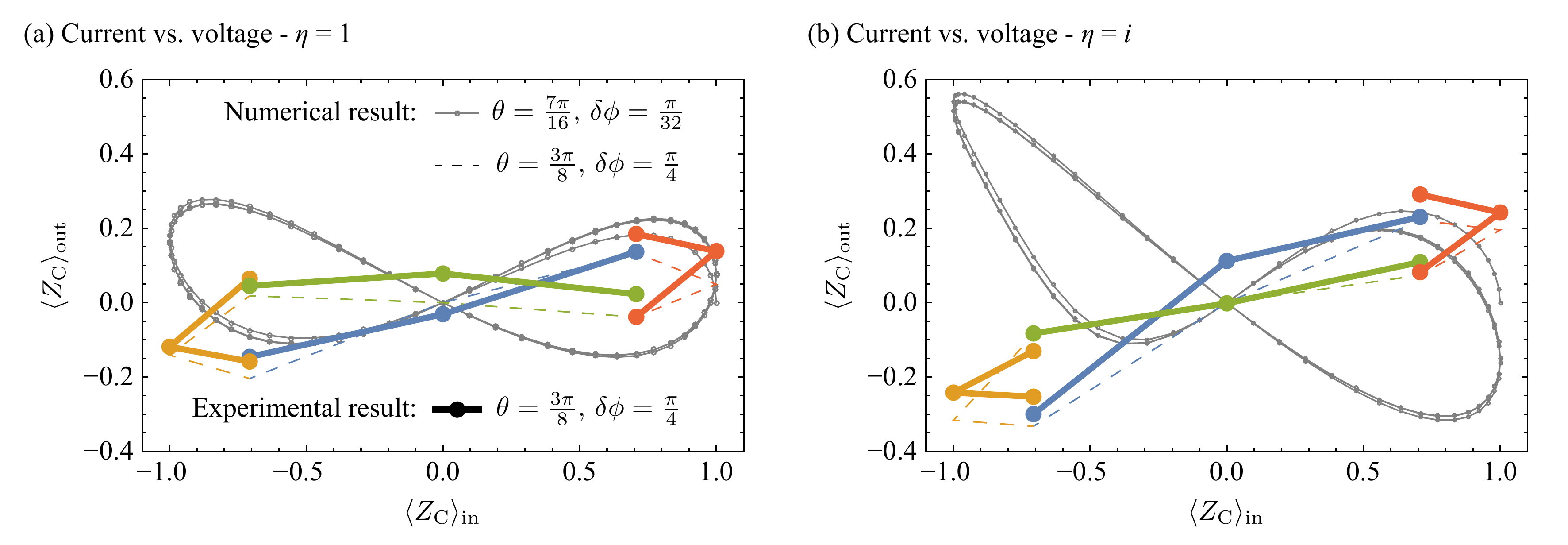}
\caption{
Hysteresis loops of memristive gates. Here $\mean{Z_{\rm C}}_{\rm in}$ and $\mean{Z_{\rm C}}_{\rm out}$ represent the voltage and current, respectively. Details are in Appendix~\ref{app:Hysteresis}. 
}
\label{fig:memristor}
\end{figure*}

An artificial neural network with three layers is proposed as an example of the neuromorphic system based on memristive quantum gates, as shown in Fig.~\ref{fig:neural_network}(d). Neurons in the input and hidden layers are qubits, and neurons in the output layer are classical bits. Two quantum layers are wired up by memristive gates, and output bits are measurement outcomes of hidden-layer qubits. Compared with the general quantum neural network~\cite{Beer2020}, the number of variational parameters is significantly reduced with respect to the number of neurons and synapses. Each connection between an input neuron and a hidden-layer neuron is characterised by two variational parameters (i.e.~weights), and each connection to an output neuron is characterised by only one parameter. We prove that such a three-layer memristive neural network is as powerful as a universal quantum computer~\cite{Deutsch1985} up to a polynomial overhead. The application of the neural network is demonstrate in quantum state classification tasks~\cite{Farhi2018, Grant2018, Schuld2020, Gao2018}. 

All the results are demonstrated with numerical simulations using QuESTlink~\cite{Jones2019} and experiments on the quantum computer {\it ibmq{\textunderscore}vigo}. An example circuit realisation of memristive quantum gates is given in Fig.~\ref{fig:neural_network}(c). Alternative circuits are used in experiments for minimising the impact of errors. Details of numerical simulations and experiments are in Appendix. 

\section{Memristive quantum gates}

To find quantum gates with the memristive properties, we introduce a simplified picture of the memristor, which is different from actual memristive devices~\cite{Strukov2008}. When we send the input current to the memristor, the current is transmitted or reflected depending on the state of memristor, and the state of memristor evolves depending on the input current. If the input current is from A to B~[see Fig.~\ref{fig:neural_network}(b)], the resistance of the memristor decreases. If the input current is from B to A, the resistance increases. We use one qubit to represent the current state: $\ket{0}_{\rm C}$ and $\ket{1}_{\rm C}$ denote currents from A to B and from B to A, respectively. We use another qubit to represent the resistance state: $\ket{0}_{\rm R}$ and $\ket{1}_{\rm R}$ denote transmission and reflection, respectively. In the extreme case, the resistance state can be completely flipped in one shot, then the memristor is the transformation $\ket{0}_{\rm C}\otimes\ket{0}_{\rm R} \rightarrow \ket{0}_{\rm C}\otimes\ket{0}_{\rm R}$, $\ket{0}_{\rm C}\otimes\ket{1}_{\rm R} \rightarrow \ket{1}_{\rm C}\otimes\ket{0}_{\rm R}$, $\ket{1}_{\rm C}\otimes\ket{0}_{\rm R} \rightarrow \ket{1}_{\rm C}\otimes\ket{1}_{\rm R}$ and $\ket{1}_{\rm C}\otimes\ket{1}_{\rm R} \rightarrow \ket{0}_{\rm C}\otimes\ket{1}_{\rm R}$. The key point is that input states and output states of this transformation are both orthogonal. Therefore, it can be a unitary transformation, i.e.~a quantum gate. 

Now, we consider the general case that the resistance state is rotated by a finite angle of $\pi - 2\theta$ when it is not saturated. The corresponding unitary transformation reads 
\begin{eqnarray}
M_\theta = \left( \begin{array}{cccc}
1 & 0 & 0 & 0 \\
0 & 0 & 0 & e^{i\theta} \\
0 & \cos\theta & i\sin\theta & 0 \\
0 & ie^{-i\theta}\sin\theta & e^{-i\theta}\cos\theta & 0
\end{array}\right),
\end{eqnarray}
where basis vectors are sorted as $\ket{0}_{\rm C}\otimes\ket{0}_{\rm R}$, $\ket{0}_{\rm C}\otimes\ket{1}_{\rm R}$, $\ket{1}_{\rm C}\otimes\ket{0}_{\rm R}$ and $\ket{1}_{\rm C}\otimes\ket{1}_{\rm R}$. When $\theta = 0$, $M_\theta$ can flip the resistance state in one shot as in the extreme case. When $\theta$ is finite, the gate transforms the input state $\ket{0}_{\rm C}\otimes\ket{1}_{\rm R}$ into $\ket{1}_{\rm C}\otimes(\cos\theta\ket{0}_{\rm R} + ie^{-i\theta}\sin\theta\ket{1}_{\rm R})$, i.e.~the current is reflected, and the resistance state is rotated by a finite angle. It is similar for the input state $\ket{1}_{\rm C}\otimes\ket{0}_{\rm R}$. We can find that the influence of the input current on the resistance state is minimised at $\theta = \frac{\pi}{2}$. 

Many similar memristive gates can be constructed. For example, we can change the phases $e^{i\theta}$ and $e^{-i\theta}$, and the gate is still memristor-like. We choose the phases such that the gate $M_\theta$ can be used for encoding a quantum state and implementing universal quantum computing on the neural network, as we will show later. 

In some scenarios, we want to use different qubits to represent the states of two terminals A and B. For example, we use two qubits A and B to represent the voltages of two terminals. We can modify the memristive gate by taking $\ket{0}_{\rm C} = \ket{1}_{\rm A}\otimes\ket{0}_{\rm B}$ and $\ket{1}_{\rm C} = \ket{0}_{\rm A}\otimes\ket{1}_{\rm B}$. Then, a three-qubit memristive gate is $\widetilde{M}_\theta = M_\theta \oplus \openone_4$, where $\openone_4$ is the four-dimensional identity matrix acting on the subspace of  $\ket{0}_{\rm A}\otimes\ket{0}_{\rm B}\otimes\ket{\mu}_{\rm R}$ and $\ket{1}_{\rm A}\otimes\ket{1}_{\rm B}\otimes\ket{\mu}_{\rm R}$, i.e.~the state of memristor does not change when two terminals have the same voltage. Qubits A and B can also be used to represent the spike timings of two neurons when the resistance qubit mimics the synapse. Memristive quantum gates for multi-state current and resistance can be constructed in a similar way. In this paper, we focus on the two-qubit gate for simplicity. 

\begin{figure}[tbp]
\centering
\includegraphics[width=1\linewidth]{\figpath 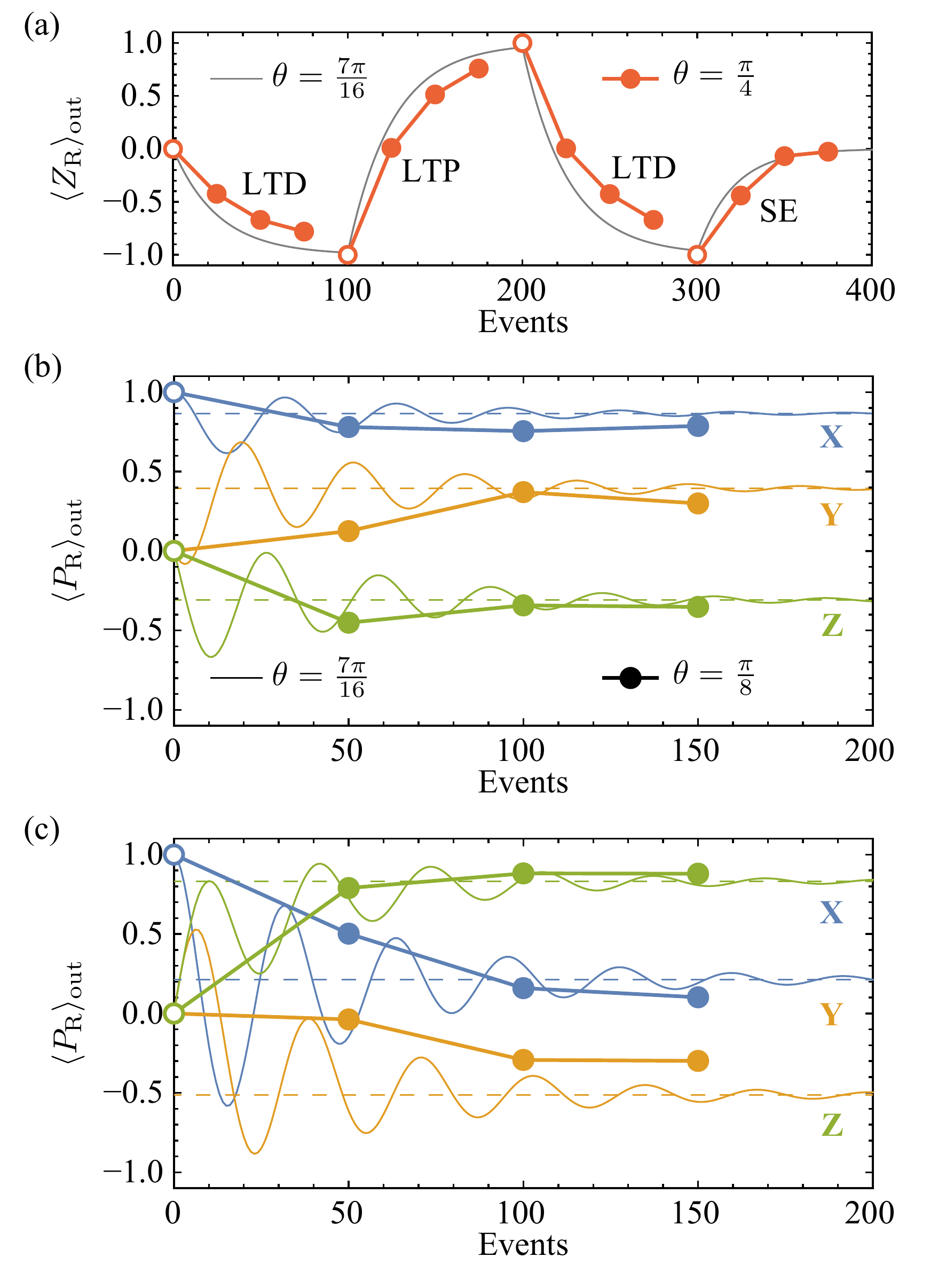}
\caption{
(a) Classical and (b,c) quantum long-term plasticity based on memristive gates. Thin solid curves represent numerical results, and filled circles represent experimental results. Empty circles denote initial values in the experiments. In (b) and (c), we take the same values of the parameter $\theta$. Dashed horizontal lines denote values in the input state of current qubits, i.e.~the steady state. $P = X,Y,Z$ are Pauli operators. The quantum state is successfully encoded when three $\mean{P_{\rm R}}_{\rm out}$ converge to dashed lines. See Appendix~\ref{app:LTP} for details. 
}
\label{fig:LTP_encoding}
\end{figure}

\section{Memristive behaviour}

Let  $\rho_{\rm C}$ and $\rho_{\rm R}$ be input states of the current qubit and resistance qubit, respectively. Then the output state after the memristive gate is $\rho_{\rm out} = M_\theta \rho_{\rm C}\otimes \rho_{\rm R} M_\theta^\dag$. If we consider mean values of the Pauli operator $Z$, we can find Ohm's law of the memristive gate, i.e.~$\mean{Z_{\rm C}}_{\rm out} = \mean{Z_{\rm R}}_{\rm in} \mean{Z_{\rm C}}_{\rm in}$, where $\mean{Z_{\rm C}}_{\rm in} = \Tr(Z\rho_{\rm C})$, $\mean{Z_{\rm R}}_{\rm in} = \Tr(Z\rho_{\rm R})$ and $\mean{Z_{\rm C}}_{\rm out} = \Tr(Z\otimes I \rho_{\rm out})$ play the roles of voltage, conductance and current, respectively. See Appendix~\ref{app:Ohm}. Here, $I$, $X$, $Y$ and $Z$ are Pauli operators. 

To demonstrate the hysteretic behaviour, we let the resistance qubit interact with a sequence of current qubits in the input states $\rho_{\rm C}^{(0)},\rho_{\rm C}^{(1)},\ldots,\rho_{\rm C}^{(t)},\ldots$ one by one through memristive gates. These states have an oscillatory observable $\mean{Z_{\rm C}}_{\rm in}(t)$, and $t$ is the label of the time. Driven by current qubits, the resistance state (i.e. the conductance) evolves with $t$, which results in the hysteretic behaviour. The $\mean{Z_{\rm C}}_{\rm out}$-versus-$\mean{Z_{\rm C}}_{\rm in}$ (i.e. current-versus-voltage) hysteresis loops are shown in Fig.~\ref{fig:memristor}. We take $\rho_{\rm C}^{(t)} = \ketbra{\psi(t)}{\psi(t)}$ as pure states, where $\ket{\psi(t)} = \cos\frac{\delta\phi t}{2}\ket{0} + \eta \sin\frac{\delta\phi t}{2}\ket{1}$, and $\eta = 1,i$ in (a) and (b), respectively. In both cases, $\mean{Z_{\rm C}}_{\rm in}(t) = \cos(\delta\phi t)$. However, the phases of quantum states are different. As a result, hysteresis loops have different shapes. 

\section{Long-term plasticity}

In STDP, causal events increase the strength of a synapse, and acausal events decrease the strength, which are called LTP and LTD, respectively. LTP and LTD can be mimicked using the memristor~\cite{Serb2016}. In the memristive gate, the resistance state evolves driven by the current qubit. The output state of the resistance qubit is $\mathcal{M}_{\theta,\rho_{\rm C}}(\rho_{\rm R}) = \Tr_{\rm C}(\rho_{\rm out})$, where $\Tr_{\rm C}$ denotes the partial trace on the current qubit, and $\mathcal{M}_{\theta,\rho_{\rm C}}$ is a completely positive map depending on $\theta$ and the input state $\rho_{\rm C}$ of the current qubit. The steady state of the map is $\rho_{s} = \frac{1}{2}(I + \mean{Z_{\rm C}}_{\rm in}Z)$ (see Appendix~\ref{app:SteadyStates}). Therefore, after the interaction with a sequence of current qubits in the same input state, the conductance of memristor converges to $\mean{Z_{\rm R}}_{\rm s} = \mean{Z_{\rm C}}_{\rm in}$, i.e.~the classical information of the current qubit is encoded into the resistance qubit. 

To demonstrate LTP and LTD phenomena mimicked using memristive gates, we take $\rho_{\rm C} = \ketbra{0}{0}$ and $\rho_{\rm C} = \ketbra{1}{1}$ to represent causal events in LTP and acausal events in LTD, respectively. We also take $\rho_{\rm C} = \ketbra{+}{+}$ to represent stochastic events (SE) without a definite casual order, where $\ket{\pm} = \frac{1}{\sqrt{2}}(\ket{0}\pm\ket{1})$. The results of numerical simulation and experiments are shown in Fig.~\ref{fig:LTP_encoding}(a). In three-qubit memristive gates, we can use qubits A and B to represent spike timings of two neurons, which will lead to similar results. 

\section{Encoding quantum states}

Memristive gates can also encode quantum information into the resistance qubit. In LTP and LTD processes, only the classical information is encoded because the phase information is not preserved. The current qubit is flipped or not flipped depending on the resistance state. Therefore two qubits are correlated in the $Z$ direction in the output state, which damages the phase information. To restore the phase, we can measure the output current qubit in the $X$ basis and adjust the phase of the resistance qubit: the identity gate $I$ or phase gate $Z$ on the resistance qubit is performed if the measurement outcome is $\ket{+}$ or $\ket{-}$, respectively. Accordingly, the map on the resistance qubit reads  $\mathcal{M}'_{\theta,\rho_{\rm C}}(\rho_{\rm R}) = \Tr_{\rm C}(K_+ \rho_{\rm out} K_+) + \Tr_{\rm C}(K_- \rho_{\rm out} K_-)$, where $K_\eta = \ketbra{\eta}{\eta}\otimes Z^{\frac{1}{2}-\eta\frac{1}{2}}$. The steady state of the map is $\rho'_{\rm s} = \rho_{\rm C}$ (see Appendix~\ref{app:SteadyStates}). Therefore, after the interaction with a sequence of current qubits in the same input state, the resistance state converges to $\rho_{\rm C}$, i.e.~the quantum information is encoded. 

The  quantum state encoding is demonstrated in Figs.~\ref{fig:LTP_encoding}(b)~and~(c). The input state $\rho_{\rm C}$ is $e^{-i\frac{7\pi}{22}Z} e^{-i\frac{3\pi}{10}X} \ket{0}$ in (b) and $e^{-i\frac{\pi}{16}Z} e^{-i\frac{3\pi}{32}X} \ket{0}$ in (c). In the two experiments, the encoding fidelity reaches $97.672\%$ and $97.638\%$ after three memristive gates in (b) and (c), respectively. 

\begin{figure}[tbp]
\centering
\includegraphics[width=1\linewidth]{\figpath 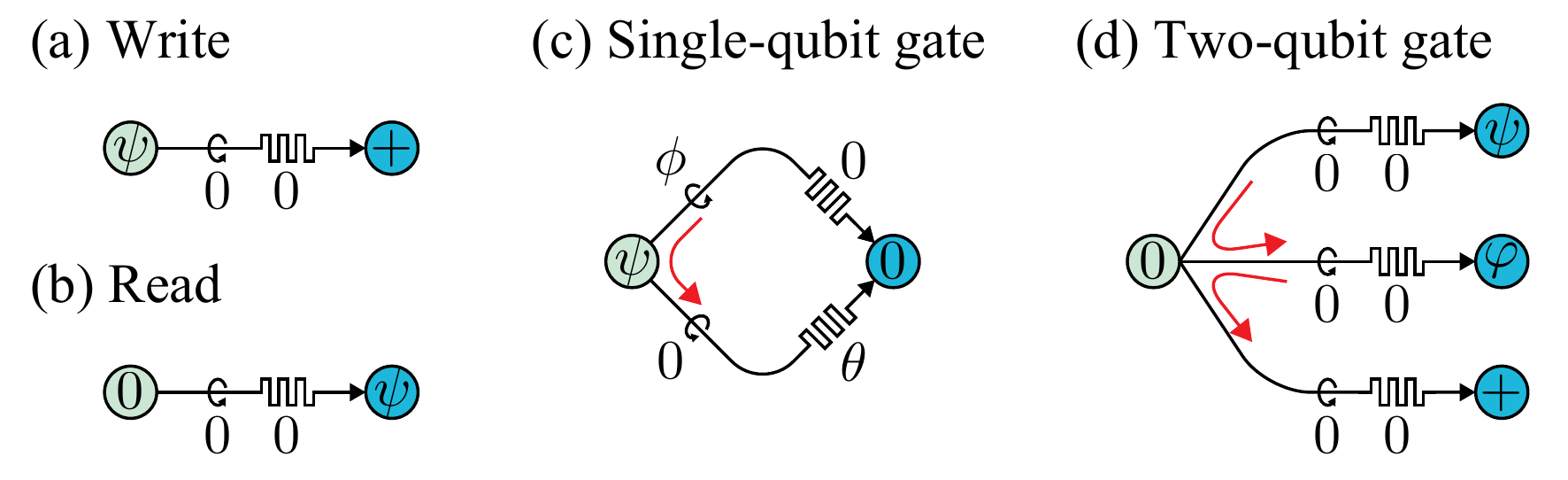}
\caption{
Universal quantum computing operations on the neural network. Red arrows denote the time sequence. 
}
\label{fig:gates}
\end{figure}

\section{Artificial neural network}

The neural network in Fig.~\ref{fig:neural_network}(d) has three layers. The input layer and hidden layer are formed by $M$ current qubits and $N$ resistance qubits, respectively. Each connection between the two quantum layers has three labels $(i,a,b)$ and two parameters $(\phi_i,\theta_i)$: The $i$-th connection is a composite gate $M_{\theta_i} e^{-i\frac{\phi_i}{2}Y}\otimes I$ on the $a$-th current qubit and $b$-th resistance qubit. Here, the $Y$-axis rotation is on the current qubit. We remark that these connections are time-ordered according to $i$ because quantum gates are non-commutative. The output layer is formed by $N$ classical bits. Each resistance qubit and the corresponding classical bit has a connection with only one parameter $\phi_j$: After a $Y$-axis rotation $e^{-i\frac{\phi_j}{2}Y}$, the resistance qubit is measured in the $Z$ basis, and the outcome is the classical bit. 

\section{Universal quantum computing}

To implement the universal quantum computing on the memristive artificial neural network, we initialise input (current) and hidden-layer (resistance) qubits in states $\ket{0}$ and $\ket{+}$, respectively. We can think of that resistance qubits form the register of quantum data, and current qubits conduct the computing. (i) A current qubit can write/read the quantum state of a resistance qubit by taking $\phi = \theta = 0$, as shown in Fig.~\ref{fig:gates}(a), corresponding to transformations $M_0\ket{\psi}\otimes\ket{+} = \ket{+}\otimes\ket{\psi}$ and $M_0\ket{0}\otimes\ket{\psi} = \ket{\psi}\otimes\ket{0}$, respectively. (ii) To perform a single-qubit gate, we let a current qubit carry the qubit state $\ket{\psi}$ and prepare a resistance qubit in the state $\ket{0}$ by using write/read operations. Then, by visiting the resistance qubit twice with parameters shown in Fig.~\ref{fig:gates}(b), we obtain the transform $M_\theta M_0 e^{-i\frac{\phi}{2}Y}\otimes I\ket{\psi}\otimes\ket{0} = I\otimes e^{-i\frac{\theta}{2}Z} e^{-i\frac{\phi}{2}Y} \ket{0}\otimes\ket{\psi}$, which is a universal single-qubit gate. (iii) To perform a two-qubit gate on two resistance qubits, we use a current qubit to read the state of the first qubit $\psi$ and let it interact with the second qubit $\varphi$ [see Fig.~\ref{fig:gates}(c)]. The output current state is written into the third resistance qubit. In this way, a controlled-NOT gate $\Lambda_X$ is performed. The corresponding transformation on three resistance qubits is $\ket{\Psi}_{1,2}\otimes\ket{+}_3 \rightarrow \ket{0}_1\otimes \Lambda_X \ket{\Psi}_{2,3}$, where $\ket{\Psi}$ is the input two-qubit state, and the second qubit is the control qubit in $\Lambda_X$. The universal single-qubit gate and controlled-NOT gate form a universal gate set~\cite{Nielsen2010}. 

Each controlled-NOT gate consumes one current qubit and one resistance qubit. The single-qubit gate can be implemented under the restriction that each current qubit can only visit a resistance qubit at most once. See Appendix~\ref{app:universalQC} for details. Under this restriction, each single-qubit gate consumes three current qubits and two resistance qubits. Therefore, the overhead cost is polynomial. 

\begin{figure}[tbp]
\centering
\includegraphics[width=1\linewidth]{\figpath 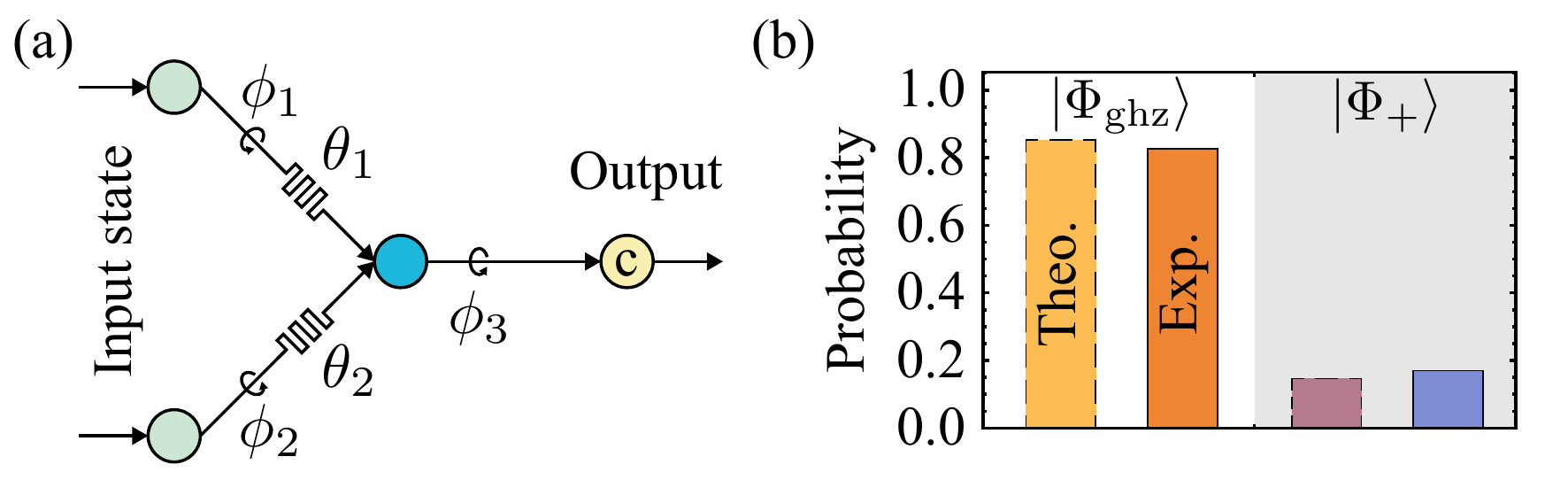}
\caption{
(a) Neural network for the classification of two-qubit states. (b) Probability of the output bit $0$ given optimal parameters. Dashed and solid boxes represent the theoretical and experimental results, respectively. 
}
\label{fig:Ising}
\end{figure}

\section{Quantum state classification}

Now, we use the memristive neural network for the quantum state classification~\cite{Farhi2018, Grant2018, Schuld2020, Gao2018}. Input qubits are prepared in one of quantum states to be classified $\ket{\Phi_k}$. Hidden-layer qubits are initialised in the state $\ket{+}$. The probability distribution of output classical bits $\boldsymbol{\mu}$ is $p_{\boldsymbol{\phi},\boldsymbol{\theta}}(\boldsymbol{\mu}\vert\Phi_k)$ given the input state $\ket{\Phi_k}$, where $\boldsymbol{\phi}$ and $\boldsymbol{\theta}$ are parameters of the neural network. We find the optimal parameters by maximising $\overline{D} = \sum_{k\neq k'} D(p_{\boldsymbol{\phi},\boldsymbol{\theta}}(\bullet\vert\Phi_k),p_{\boldsymbol{\phi},\boldsymbol{\theta}}(\bullet\vert\Phi_{k'}))$. Here, $D$ is the trace distance between two distributions~\cite{Nielsen2010}, which characterises how well two states can be distinguished according to the output $\boldsymbol{\mu}$. 

Two examples are implemented. First, we use a network with two neurons in each layer, i.e.~$M=N=2$ to classify four Bell states. Because Bell states are orthogonal, they are completely distinguishable, which can be achieved by the neural network. Second, we use a network with $M=N=5$ to classify two five-qubit ground states of the quantum Ising model in ferromagnetic and paramagnetic phases~\cite{Sachdev1999}, i.e.~the Greenberger-Horne-Zeilinger state $\ket{\Phi_{\rm ghz}} = \frac{1}{\sqrt{2}}(\ket{0}^{\otimes M}+\ket{1}^{\otimes M})$ and the product state $\ket{\Phi_+} = \ket{+}^{\otimes M}$. These two states are not orthogonal. We find that the maximum distance given by the neural network can reach the quantum upper bound, i.e.~the trace distance between two quantum states~\cite{Nielsen2010}. If we turn off parameters $\boldsymbol{\phi}$ by setting all $\phi$ to zero, only memristive gates are used in the classification. In this case, the distance can reach $0.94792$, which is lower than the upper bound $0.96824$ but is still above the classical value $0.9375$, i.e.~the distance given by a direct measurement in the $Z$ basis on each qubit. Numerical data of the optimisation computing are in Appendix~\ref{app:QSC}. For the experimental implementation, we use the network shown in Fig.~\ref{fig:Ising}(a) to classify two-qubit ground states. In the numerical simulation, the distance can reach the theoretical upper bound $0.70711$, which is reduced to $0.65673$ (but still higher than the classical value $0.5$) in the experiment using optimal parameters. The corresponding distributions are shown in Fig.~\ref{fig:Ising}(b). 

\section{Discussion}

We have demonstrated that memristive quantum gates can mimic memristors and synapses, which are essential building blocks of neuromorphic computing. These gates are unitary transformations that are feasible in many physical systems~\cite{Nielsen2010}. Memristive gates are fully quantum compared with the memristance involving the weak measurement and dissipation in quantum systems~\cite{Pfeiffer2016, Salmilehto2017, Sanz2018, GonzalezRaya2020, Maier2015, Li2017PRB}. The experiments are implemented using universal gates on a circuit-based quantum computer {\it ibmq{\textunderscore}vigo}. By engineering the interaction between qubits, it is also possible to realise a memristive gate directly in the time evolution. Synapses based on memristive gates can encode the quantum state in a way similar to the long-term plasticity, therefore, are capable of processing quantum information. We have demonstrated the supervised quantum state classification on the memristive neural network, which can also be used for the unsupervised learning~\cite{Serb2016}. These results pave the way towards the neuromorphic system in the quantum regime, i.e.~a brain-inspired quantum computer. 

\begin{acknowledgments}
This work is supported by National Natural Science Foundation of China (Grant No. 11875050) and NSAF (Grant No. U1930403). YL thanks Tyson Jones for help on using QuESTlink. 
\end{acknowledgments}

\appendix

\setcounter{figure}{0}
\setcounter{table}{0}
\renewcommand\thefigure{S\arabic{figure}}
\renewcommand\thetable{S\arabic{table}}

\section{Numerical simulation and experiment}

We implement numerical simulations using QuESTlink, which is a library based on the framework of Quantum Exact Simulation Toolkit (QuEST). We perform experiments on {\it ibmq{\textunderscore}vigo} via IBM Quantum Experience. 

The superconducting quantum computer {\it ibmq{\textunderscore}vigo} has five qubits. Two-qubit gates are available on nearest neighbouring qubits (0,1), (1,2), (1,3) and (3,4). Only qubits 0,1,2,3 are used in the experiments. In the calibration data from IBM Quantum Experience on 25 Feb 2020, single-qubit-gate error rates are from 0.03\% to 0.07\%, and two-qubit gate error rates are from 0.68\% to 1.18\%, depending on the qubits. We performed experiments on 25-27 Feb 2020. Each circuit runs for 8192 shots in experiments. 

In all the experiments, circuits are altered from Fig.~1(c) and optimised for minimising the impact of errors on {\it ibmq{\textunderscore}vigo}. In the hysteresis, LTP, LTD and quantum state encoding experiments, the qubit 1 is the resistance qubit, and qubits 0,2,3 are current qubits. In quantum state encoding experiments, we replace the measurement and feedback phase gate with a controlled-NOT gate, and they result in the same effect on the resistance qubit when gates are perfect. In the quantum state classification experiment, qubits 0,1,2 are used, and the roles (resistance or current) of qubits change in the circuit for minimising the number of two-qubit gates. More details will be given in the following sections. 

\section{Ohm's law}
\label{app:Ohm}

Consider the transformation of the operator $Z\otimes I$, we have 
\begin{eqnarray}
M_\theta^\dag Z\otimes I M_\theta = Z\otimes Z.
\end{eqnarray}
Therefore, 
\begin{eqnarray}
\mean{Z_{\rm C}}_{\rm out} &=& \Tr(Z\otimes I \rho_{\rm out}) \notag \\
&=& \Tr(Z\otimes Z \rho_{\rm C}\otimes \rho_{\rm R}) = \mean{Z_{\rm R}}_{\rm in} \mean{Z_{\rm C}}_{\rm in}.
\end{eqnarray}

\begin{figure*}[tbp]
\centering
\includegraphics[width=1\linewidth]{\figpath 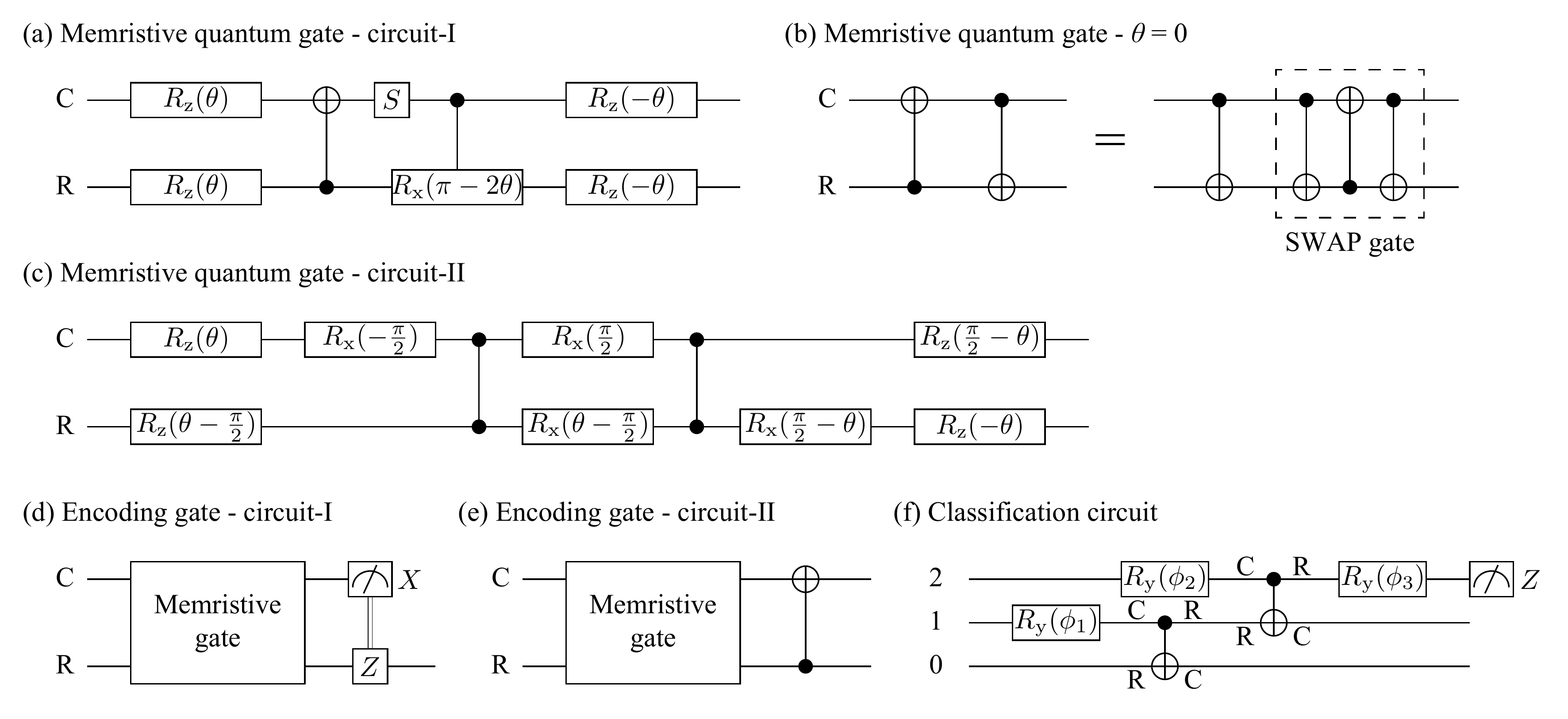}
\caption{
(a) Detailed display of the circuit of the memristive gate $M_{\theta}$ in Fig.~1(c). (b) Circuit of the memristive gate $M_{0}$. (c) Circuit of the memristive gate $M_{\theta}$ optimised for the implementation on {\it ibmq{\textunderscore}vigo}. (d) Circuit of the modified memristive gate with a measurement on the current qubit and a feedback gate on the resistance qubit. (e) Circuit of the modified memristive gate with an additional controlled-NOT gate, which is used in experiments on {\it ibmq{\textunderscore}vigo}. (f) Circuit of the three-neuron neural network experiment for the state classification implemented on {\it ibmq{\textunderscore}vigo}. Here, $R_{x}(\phi) = e^{-i\frac{\phi}{2}X}$, $R_{y}(\phi) = e^{-i\frac{\phi}{2}Y}$ and $R_{z}(\phi) = e^{-i\frac{\phi}{2}Z}$. 
}
\label{fig:circuits}
\end{figure*}

\section{Hysteresis loops}
\label{app:Hysteresis}

In the numerical simulations, the resistance qubit is initialised in the state $\rho_{\rm R}^{(0)} = \ketbra{+}{+}$, where $\ket{+} = \frac{1}{\sqrt{2}}(\ket{0}+\ket{1})$. With this initial state, we compute the output states of the first gate, $\rho_{\rm out}^{(0)} = M_\theta\rho_{\rm C}^{(0)}\otimes\rho_{\rm R}^{(0)}M_\theta^\dag$ and $\rho_{\rm R}^{(1)} = \Tr_{\rm C}(\rho_{\rm out}^{(0)})$; with the output resistance state of the first gate, we compute the output states of the second gate, $\rho_{\rm out}^{(1)} = M_\theta\rho_{\rm C}^{(1)}\otimes\rho_{\rm R}^{(1)}M_\theta^\dag$ and $\rho_{\rm R}^{(2)} = \Tr_{\rm C}(\rho_{\rm out}^{(1)})$; and so on. In this way, we can obtain output states of each gate. Then, at the time $t$, the voltage is $\mean{Z_{\rm C}}_{\rm in}(t) = \Tr(Z\rho_{\rm C}^{(t)})$, the output current is $\mean{Z_{\rm C}}_{\rm out}(t) = \Tr(Z\otimes I\rho_{\rm out}^{(t)})$, and the output conductance is $\mean{Z_{\rm R}}_{\rm out}(t) = \Tr(I\otimes Z\rho_{\rm out}^{(t)})$. 

In Fig.~2, small gray circles represent the numerical data of $(\mean{Z_{\rm C}}_{\rm in}(t),\mean{Z_{\rm C}}_{\rm out}(t))$ with $\theta = \frac{7\pi}{16}$ and $\delta\phi = \frac{\pi}{32}$, where $t = 0,1,\ldots,\frac{20\pi}{\delta\phi}-1$. Dashed lines represent the numerical data with $\theta = \frac{3\pi}{8}$ and $\delta\phi = \frac{\pi}{4}$. For dashed lines, the numerical simulations are implemented for $t = 0,1,\ldots,\frac{20\pi}{\delta\phi}+1$, however, only the last cycle is plotted, i.e.~$t = \frac{20\pi}{\delta\phi}-7,\frac{20\pi}{\delta\phi}-6,\ldots,\frac{20\pi}{\delta\phi}+1$. The blue dashed lines represent $t = \frac{20\pi}{\delta\phi}-7,\frac{20\pi}{\delta\phi}-6,\frac{20\pi}{\delta\phi}-5$; the yellow dashed lines represent $t = \frac{20\pi}{\delta\phi}-5,\frac{20\pi}{\delta\phi}-4,\frac{20\pi}{\delta\phi}-3$; the green dashed lines represent  $t = \frac{20\pi}{\delta\phi}-3,\frac{20\pi}{\delta\phi}-2,\frac{20\pi}{\delta\phi}-1$; and the orange dashed lines represent  $t = \frac{20\pi}{\delta\phi}-1,\frac{20\pi}{\delta\phi},\frac{20\pi}{\delta\phi}+1$. 

On {\it ibmq{\textunderscore}vigo}, a qubit has direct gate coupling with at most three other qubits. Therefore, we implement the memristive gates between the resistance qubit and at most three current qubits. To demonstrate a full cycle of each hysteresis loop, we divide the cycle into four segments, i.e.~four experiments, according to the four segments of the dashed lines. In the experiments, we take $\theta = \frac{3\pi}{8}$ and $\delta\phi = \frac{\pi}{4}$ as the same as in numerical simulations of the dashed lines. For the segment started at $t = s$, we prepare the resistance qubit in the numerically-computed output state $\rho_{\rm R}^{(s)}$, and then we let the resistance qubit interact with three current qubits prepared in states $\rho_{\rm C}^{(s)}$, $\rho_{\rm C}^{(s+1)}$ and $\rho_{\rm C}^{(s+2)}$ one by one. For the first segment (large blue circles), $s = \frac{20\pi}{\delta\phi}-7$; for the second segment (large yellow circles), $s = \frac{20\pi}{\delta\phi}-5$; for the third segment (large green circles), $s = \frac{20\pi}{\delta\phi}-3$; and for the forth segment (large orange circles), $s = \frac{20\pi}{\delta\phi}-1$. Data $\mean{Z_{\rm C}}_{\rm out}(t)$ are measured in the experiments, and $(\mean{Z_{\rm C}}_{\rm in}(t),\mean{Z_{\rm C}}_{\rm out}(t))$ are plotted as large circles in Fig.~2. If quantum gates are ideal, experimental data should be consistent with dashed lines. The difference is caused by the noise on {\it ibmq{\textunderscore}vigo}. 

\begin{figure*}[tbp]
\centering
\includegraphics[width=1\linewidth]{\figpath 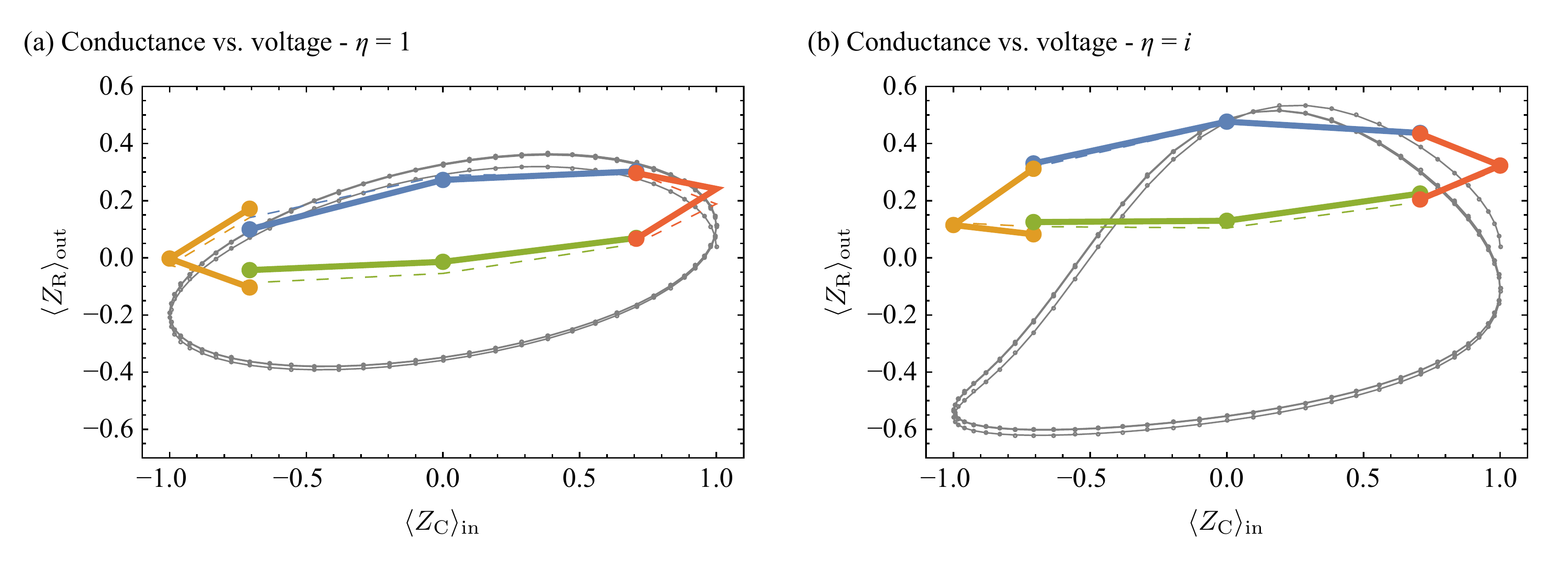}
\caption{
Hysteresis loops of $(\mean{Z_{\rm C}}_{\rm in}(t),\mean{Z_{\rm R}}_{\rm out}(t))$. Here $\mean{Z_{\rm C}}_{\rm in}$ and $\mean{Z_{\rm R}}_{\rm out}$ represent the voltage and conductance, respectively. 
}
\label{fig:memristorII}
\end{figure*}

In the experiments, we decompose the memristive gate into elementary gates as shown in Fig.~\ref{fig:circuits}(c). Data of $(\mean{Z_{\rm C}}_{\rm in}(t),\mean{Z_{\rm R}}_{\rm out}(t))$ are shown in Fig.~\ref{fig:memristorII}. 

\section{Steady states of maps}
\label{app:SteadyStates}

We use the Pauli transfer matrix representation. The input state of the current qubit is $\rho_{\rm C} = \frac{1}{2}(I + \rho_{\rm C}^X X + \rho_{\rm C}^Y Y + \rho_{\rm C}^Z Z)$, and the input state of the resistance qubit is $\rho_{\rm R} = \frac{1}{2}(I + \rho_{\rm R}^X X + \rho_{\rm R}^Y Y + \rho_{\rm R}^Z Z)$, where $I$, $X$, $Y$ and $Z$ are Pauli operators. The Pauli transfer matrix of the memristive-gate maps $\mathcal{M}_{\theta,\rho_{\rm C}}$ and $\mathcal{M}'_{\theta,\rho_{\rm C}}$ are 
\begin{eqnarray}
M_{\theta,\rho_{\rm C}} = \left(\begin{array}{cc}
1 & 0 \\
k & E
\end{array}\right)~{\rm and}~
M'_{\theta,\rho_{\rm C}} = \left(\begin{array}{cc}
1 & 0 \\
k' & E'
\end{array}\right),
\end{eqnarray}
respectively, where 
\begin{eqnarray}
k = \rho_{\rm C}^Z \left(\begin{array}{c}
\cos\theta \sin^2\theta \\
\cos^2\theta \sin\theta \\
\cos^2\theta
\end{array}\right),~~
k' = \cos^2\theta \left(\begin{array}{c}
\rho_{\rm C}^X \\
\rho_{\rm C}^Y \\
\rho_{\rm C}^Z
\end{array}\right),
\end{eqnarray}
and 
\begin{widetext}
\begin{eqnarray}
E = \left(\begin{array}{ccc}
\rho_{\rm C}^X\cos\theta - \rho_{\rm C}^Y \sin^3\theta & - \rho_{\rm C}^X \cos^2\theta \sin\theta & - \cos\theta \sin^2\theta \\
\rho_{\rm C}^Y \cos^3\theta & \rho_{\rm C}^X \cos\theta \sin^2\theta  - \rho_{\rm C}^Y \sin\theta & - \cos^2\theta \sin\theta \\
- \rho_{\rm C}^Y \cos\theta \sin\theta & \rho_{\rm C}^X \cos\theta \sin\theta & \sin^2\theta
\end{array}\right),
\end{eqnarray}
\begin{eqnarray}
E' = \left(\begin{array}{ccc}
\sin^2\theta & - \rho_{\rm C}^Z \cos\theta \sin\theta & \rho_{\rm C}^Y \cos\theta \sin\theta \\
\rho_{\rm C}^Z \cos\theta \sin\theta & \sin^2\theta & - \rho_{\rm C}^X \cos\theta \sin\theta \\
- \rho_{\rm C}^Y \cos\theta \sin\theta & \rho_{\rm C}^X \cos\theta \sin\theta & \sin^2\theta
\end{array}\right).
\end{eqnarray}
\end{widetext}
We express the output state of the resistance qubit in the form $\rho_{\rm R,out} = \frac{1}{2}(I + \rho_{\rm R,out}^X X + \rho_{\rm R,out}^Y Y + \rho_{\rm R,out}^Z Z)$. If $\rho_{\rm R,out} = \mathcal{M}_{\theta,\rho_{\rm C}}(\rho_{\rm R})$, we have 
\begin{eqnarray}
\left(\begin{array}{c}
\rho_{\rm R,out}^X \\
\rho_{\rm R,out}^Y \\
\rho_{\rm R,out}^Z
\end{array}\right)
= E \left(\begin{array}{c}
\rho_{\rm R}^X \\
\rho_{\rm R}^Y \\
\rho_{\rm R}^Z
\end{array}\right) + k;
\end{eqnarray}
Similarly, if $\rho_{\rm R,out} = \mathcal{M}'_{\theta,\rho_{\rm C}}(\rho_{\rm R})$, we have 
\begin{eqnarray}
\left(\begin{array}{c}
\rho_{\rm R,out}^X \\
\rho_{\rm R,out}^Y \\
\rho_{\rm R,out}^Z
\end{array}\right)
= E' \left(\begin{array}{c}
\rho_{\rm R}^X \\
\rho_{\rm R}^Y \\
\rho_{\rm R}^Z
\end{array}\right) + k'.
\end{eqnarray}

The steady state of the map $\mathcal{M}_{\theta,\rho_{\rm C}}$ is the solution of the equation $\rho_{\rm s} = \mathcal{M}_{\theta,\rho_{\rm C}}(\rho_{\rm s})$. Express the steady state in the form $\rho_{\rm s} = \frac{1}{2}(I + \rho_{\rm s}^X X + \rho_{\rm s}^Y Y + \rho_{\rm s}^Z Z)$, the solution is 
\begin{eqnarray}
\left(\begin{array}{c}
\rho_{\rm s}^X \\
\rho_{\rm s}^Y \\
\rho_{\rm s}^Z
\end{array}\right)
= \left(\begin{array}{c}
0 \\
0 \\
\rho_{\rm C}^Z
\end{array}\right).
\end{eqnarray}
We remark that $\rho_{\rm C}^Z = \Tr(Z\rho_{\rm C}) = \mean{Z_{\rm C}}_{\rm in}$. 

Similarly, the steady state of the map $\mathcal{M}'_{\theta,\rho_{\rm C}}$ is the solution of the equation $\rho'_{\rm s} = \mathcal{M}'_{\theta,\rho_{\rm C}}(\rho'_{\rm s})$. Express the steady state in the form $\rho'_{\rm s} = \frac{1}{2}(I + \rho_{\rm s}^{\prime X} X + \rho_{\rm s}^{\prime Y} Y + \rho_{\rm s}^{\prime Z} Z)$, the solution is 
\begin{eqnarray}
\left(\begin{array}{c}
\rho_{\rm s}^{\prime X} \\
\rho_{\rm s}^{\prime Y} \\
\rho_{\rm s}^{\prime Z}
\end{array}\right)
= \left(\begin{array}{c}
\rho_{\rm C}^X \\
\rho_{\rm C}^Y \\
\rho_{\rm C}^Z
\end{array}\right).
\end{eqnarray}
Therefore, $\rho'_{\rm s} = \rho_{\rm C}$. 

\begin{figure}[tbp]
\centering
\includegraphics[width=1\linewidth]{\figpath 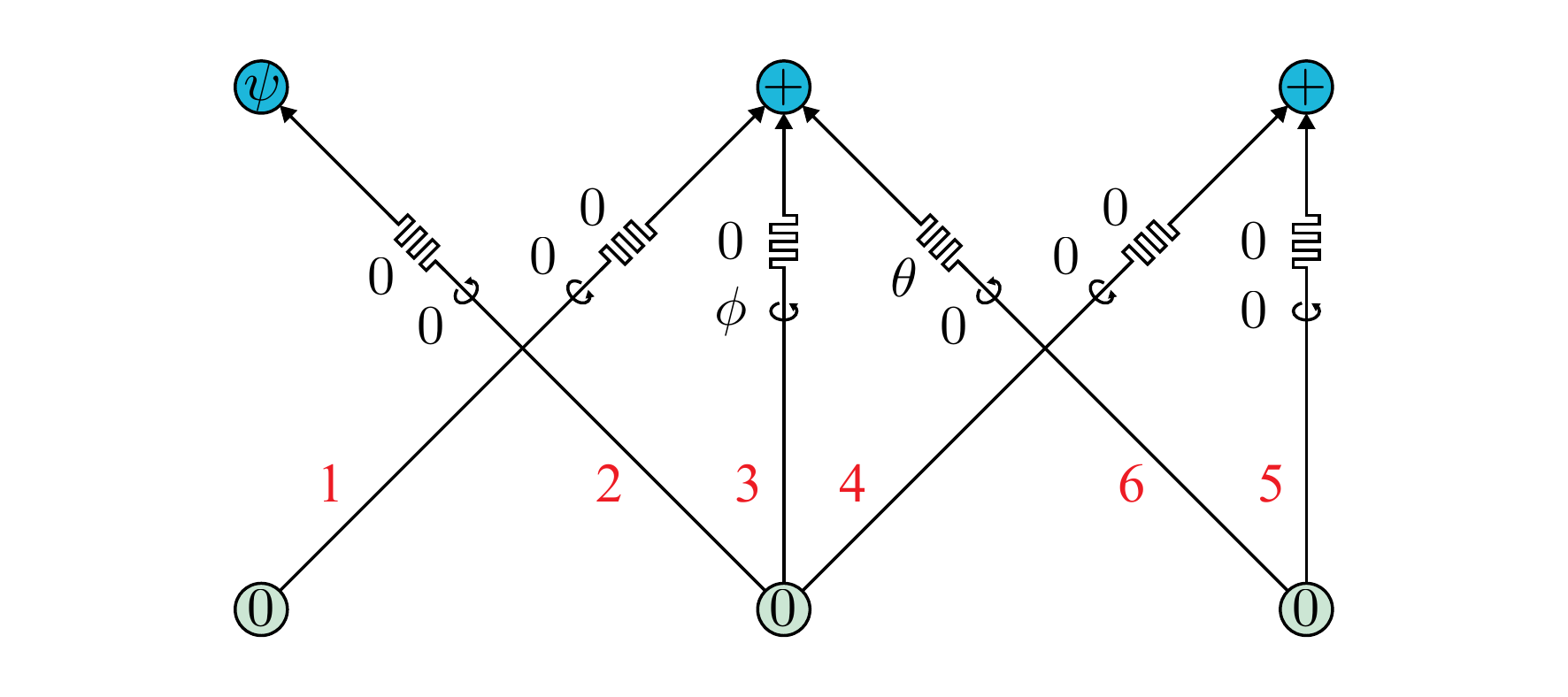}
\caption{
Single-qubit gate. Red numbers denote the time sequence. 
}
\label{fig:1q_gate}
\end{figure}

\section{LTP, LTD and quantum-state encoding}
\label{app:LTP}

In the LTP and LTD numerical simulations, we let the resistance qubit interact with a sequence of current qubits in the input states $\rho_{\rm C}^{(0)},\rho_{\rm C}^{(1)},\ldots,\rho_{\rm C}^{(t)},\ldots$ one by one through memristive gates, as the same as in hysteresis-loop simulations. We take $\rho_{\rm C}^{(t)} = \ketbra{1}{1}$ when $t = 0,1,\cdots,99$, $\rho_{\rm C}^{(t)} = \ketbra{0}{0}$ when $t = 100,101,\cdots,199$, $\rho_{\rm C}^{(t)} = \ketbra{1}{1}$ again when $t = 200,201,\cdots,299$, and $\rho_{\rm C}^{(t)} = \ketbra{+}{+}$ when $t = 300,301,\cdots,399$. The resistance qubit is initialised in the state $\rho_{\rm R}^{(0)} = \ketbra{+}{+}$. With this initial state, we compute the output states of the resistance qubit, i.e.~$\rho_{\rm R}^{(t+1)} = \mathcal{M}_{\theta,\rho_{\rm C}^{(t)}}(\rho_{\rm R}^{(t)})$, where $\theta = \frac{7\pi}{16}$. Then, $\mean{Z_{\rm R}}_{\rm out}(t) = \Tr(Z\rho_{\rm R}^{(t)})$ is computed and plotted as the thin curve in Fig.~3(a). 

Four LTP and LTD experiments are implemented on {\it ibmq{\textunderscore}vigo}, corresponding to four thick curves (with circles) in Fig.~3(a), respectively. From left to right, in the first experiment, the resistance qubit is initialised in the state $\ket{+}$, and three current qubits are initialised in the state $\ket{1}$; in the second experiment, the resistance qubit is initialised in the state $\ket{1}$, and three current qubits are initialised in the state $\ket{0}$; in the third experiment, the resistance qubit is initialised in the state $\ket{0}$, and three current qubits are initialised in the state $\ket{1}$ again; and in the forth experiment, the resistance qubit is initialised in the state $\ket{1}$, and three current qubits are initialised in the state $\ket{+}$. We let the resistance qubit interact with three current qubits one by one through the memristive gate. The memristive gate is decomposed into elementary gates as shown in Fig.~\ref{fig:circuits}(c). We take $\theta = \frac{\pi}{4}$. After each memristive gate, $\mean{Z_{\rm R}}_{\rm out}$ is measured. 

In the quantum-state encoding numerical simulations, we let the resistance qubit interact with a sequence of current qubits in the input states $\rho_{\rm C}^{(0)},\rho_{\rm C}^{(1)},\ldots,\rho_{\rm C}^{(t)},\ldots$ one by one through modified memristive gates. The circuit of the modified memristive gate (encoding gate) is shown in Fig.~\ref{fig:circuits}(d). Because we are only interested in the state of the resistance qubit, the modified memristive gate can also be realised using the circuit shown in Fig.~\ref{fig:circuits}(e). The additional controlled-NOT gate is equivalent to a phase gate on the resistance qubit depending on the phase state of the current qubit. We take $\rho_{\rm C}^{(t)} = \ketbra{\psi}{\psi}$, where $\ket{\psi} = e^{-i\frac{7\pi}{22}Z} e^{-i\frac{3\pi}{10}X} \ket{0}$ and $\ket{\psi} = e^{-i\frac{\pi}{16}Z} e^{-i\frac{3\pi}{32}X} \ket{0}$ in the two simulations. The resistance qubit is initialised in the state $\rho_{\rm R}^{(0)} = \ketbra{+}{+}$. With this initial state, we compute the output state of the resistance qubit at each time $t$, i.e.~$\rho_{\rm R}^{(t+1)} = \mathcal{M}'_{\theta,\rho_{\rm C}^{(t)}}(\rho_{\rm R}^{(t)})$, where $\theta = \frac{7\pi}{16}$. Then, the mean values of three Pauli operators $\mean{P_{\rm R}}_{\rm out}(t) = \Tr(P\rho_{\rm R}^{(t)})$ are computed and plotted as thin solid curves in Figs.~3(b)~and~(c), where $P = X,Y,Z$. 

Two quantum-state encoding experiments are implemented on {\it ibmq{\textunderscore}vigo}, corresponding to two input states $\ket{\psi} = e^{-i\frac{7\pi}{22}Z} e^{-i\frac{3\pi}{10}X} \ket{0}$ and $\ket{\psi} = e^{-i\frac{\pi}{16}Z} e^{-i\frac{3\pi}{32}X} \ket{0}$ of current qubits. In each experiment, the resistance qubit is initialised in the state $\ket{+}$, and three current qubits are initialised in the state $\ket{\psi}$. We let the resistance qubit interact with three current qubits one by one through modified memristive gates. The gate is realised using the circuit in Fig.~\ref{fig:circuits}(e), in which the memristive gate is decomposed into elementary gates as shown in Fig.~\ref{fig:circuits}(c). We take $\theta = \frac{\pi}{8}$. After each memristive gate, $\mean{P_{\rm R}}_{\rm out}$ are measured, where $P = X,Y,Z$. The data are plotted as thick curves (with circles) in Figs.~3(b)~and~(c). 

We can express states of the current qubit and resistance qubit as $\rho_{\rm C}^{(t)} = \frac{1}{2}(I + \rho_{\rm C}^X X + \rho_{\rm C}^Y Y + \rho_{\rm C}^Z Z)$ and $\rho_{\rm R}^{(t)} = \frac{1}{2}(I + \rho_{\rm R}^X X + \rho_{\rm R}^Y Y + \rho_{\rm R}^Z Z)$, respectively. Here, $\rho_{\rm C}^P = \Tr(P\rho_{\rm C}^{(t)})$ and $\rho_{\rm R}^P = \Tr(P\rho_{\rm R}^{(t)})$. Therefore, when $\Tr(P\rho_{\rm C}^{(t)}) = \Tr(P\rho_{\rm R}^{(t)})$, two states are the same. In Figs.~3(b) and (c), the dashed horizontal lines represent $\Tr(P\rho_{\rm C}^{(t)})$. Because $\rho_{\rm C}^{(t)}$ is a pure state, the fidelity $F = \sqrt{ \Tr(\rho_{\rm C}^{(t)} \rho_{\rm R}^{(t)}) } = \sqrt{ (1+\rho_{\rm C}^X\rho_{\rm R}^X+\rho_{\rm C}^Y\rho_{\rm R}^Y+\rho_{\rm C}^Z\rho_{\rm R}^Z)/2 }$. 

\begin{figure*}[tbp]
\centering
\includegraphics[width=1\linewidth]{\figpath 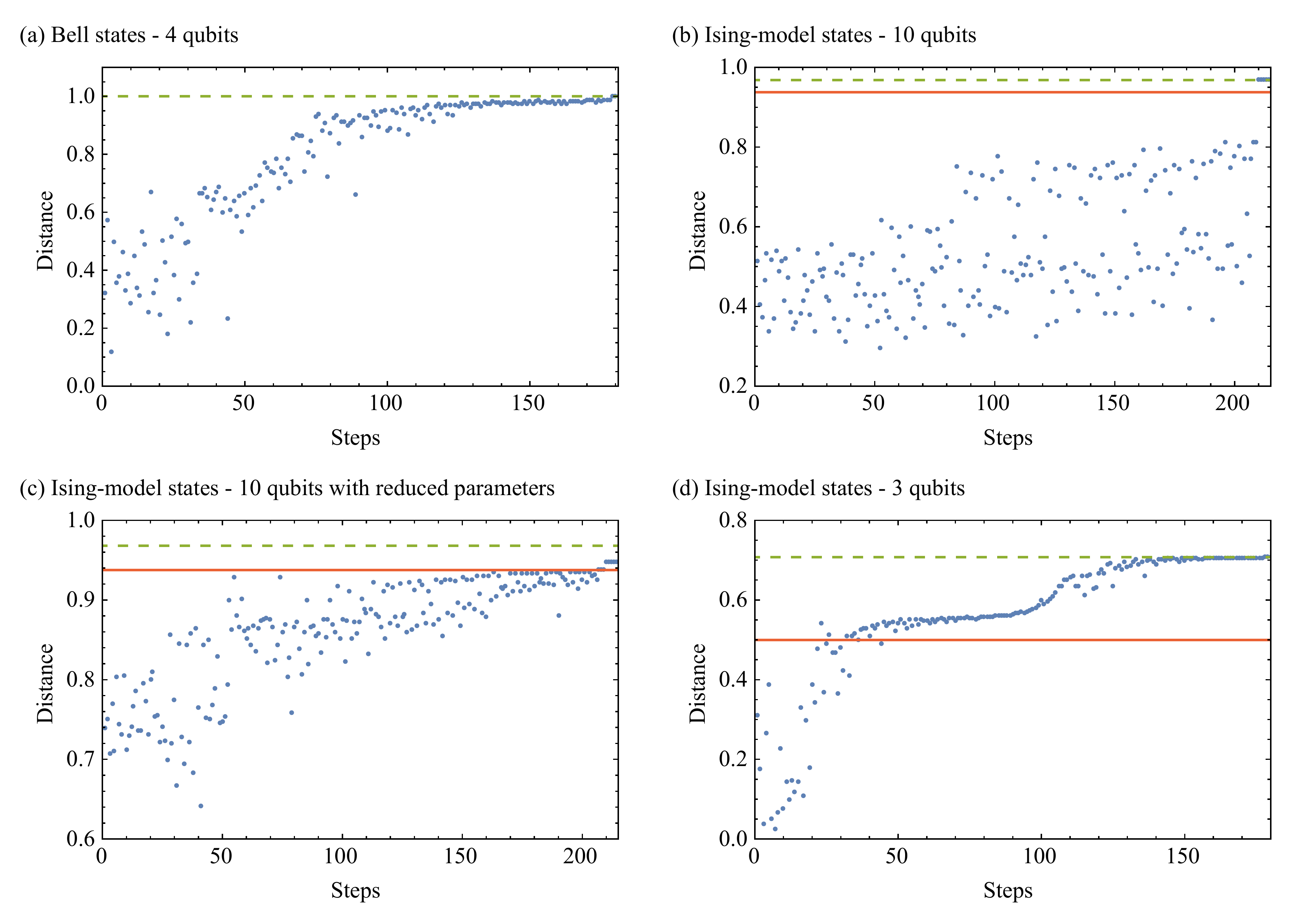}
\caption{
Values of the distance in the optimisation computing. Blue dots denote the distance returned in each step. Dashed lines denote the quantum upper bound of the distance. Solid lines denote classical values, i.e.~$D(q(\bullet\vert\Phi_{\rm ghz}),q(\bullet\vert\Phi_+))$. 
}
\label{fig:find}
\end{figure*}

\section{Universal gates}
\label{app:universalQC}

Under the restriction that each current qubit can only visit a resistance qubit once if they are connected, the single-qubit gate can be realised as shown in Fig.~\ref{fig:1q_gate}. The connection-1 prepares the second resistance qubit (from left to right) in the state $\ket{0}$. The connection-2 reads the state of the first resistance qubit $\ket{\psi}$ into the second current qubit. The connection-3 corresponds to the first visit in Fig.~4(c). Then, the connection-4 writes the output state of second current qubit into the third resistance qubit. The connection-5 reads the state of the third resistance qubit into the third current qubit. The connection-6 corresponds to the second visit in Fig.~4(c). 

To understand the controlled-NOT gate, we only need to note that the memristive gate with $\theta = 0$, i.e.~$M_0$, is equivalent to a controlled-NOT gate followed by a swap gate, as shown in Fig.~\ref{fig:circuits}(b). 

\section{Quantum state classification}
\label{app:QSC}

In the neural networks used for the quantum state classification, the input layer and the hidden layer are fully connected, and connections are sorted as follows. The first input qubit interacts with from the first to the last hidden-layer qubits one by one; then, the second input qubit interacts with from the first to the last hidden-layer qubits one by one; and so on. In other words, the $a$-th input qubit and the $b$-th hidden-layer qubit are coupled by the $i$-th $Y$-axis rotation and memristive gate, where $i = N(a-1)+b$. 

The trace distance between two distributions is 
\begin{eqnarray}
&& D(p_{\boldsymbol{\phi},\boldsymbol{\theta}}(\bullet\vert\Phi_k),p_{\boldsymbol{\phi},\boldsymbol{\theta}}(\bullet\vert\Phi_{k'})) \notag \\
&=& \frac{1}{2} \sum_{\boldsymbol{\mu}} \left\vert p_{\boldsymbol{\phi},\boldsymbol{\theta}}(\boldsymbol{\mu}\vert\Phi_k) - p_{\boldsymbol{\phi},\boldsymbol{\theta}}(\boldsymbol{\mu}\vert\Phi_{k'}) \right\vert,
\end{eqnarray}
where $\boldsymbol{\mu} = (\mu_1,\mu_2,\ldots,\mu_N)$ is a binary vector, $\mu_j$ is the value of the $j$-th output bit, i.e.~the measurement outcome of the $j$-th hidden-layer qubit, and parameter vectors are $\boldsymbol{\phi} = (\phi_1,\phi_2,\ldots,\phi_{(M+1)N})$ and $\boldsymbol{\theta} = (\theta_1,\theta_2,\ldots,\theta_{MN})$. Here, $\phi_{MN+j}$ is the parameter of the $Y$-axis rotation on the $j$-th hidden-layer qubit before the measurement. 

To distinguish four Bell states 
\begin{eqnarray}
\ket{\Phi_1} &=& \frac{1}{\sqrt{2}} (\ket{0}\otimes\ket{0}+\ket{1}\otimes\ket{1}), \\
\ket{\Phi_2} &=& \frac{1}{\sqrt{2}} (\ket{0}\otimes\ket{0}-\ket{1}\otimes\ket{1}), \\
\ket{\Phi_3} &=& \frac{1}{\sqrt{2}} (\ket{0}\otimes\ket{1}+\ket{1}\otimes\ket{0}), \\
\ket{\Phi_4} &=& \frac{1}{\sqrt{2}} (\ket{0}\otimes\ket{1}-\ket{1}\otimes\ket{0}),
\end{eqnarray}
we take $M=N=2$, i.e.~each layer has two qubits or classical bits. We find optimal parameters $\boldsymbol{\phi}$ and $\boldsymbol{\theta}$ by maximising the distance function 
\begin{eqnarray}
\overline{D}(\boldsymbol{\phi},\boldsymbol{\theta}) = \sum_{k=1}^3\sum_{k'=k+1}^{4} D(p_{\boldsymbol{\phi},\boldsymbol{\theta}}(\bullet\vert\Phi_k),p_{\boldsymbol{\phi},\boldsymbol{\theta}}(\bullet\vert\Phi_{k'})).~
\end{eqnarray}
The value of the average distance $\overline{D}/6$ is plotted in Fig.~\ref{fig:find}(a), which reaches one at the end of the optimisation. The distance $D$ is never larger than $1$, and $D = 1$ means that two states are fully distinguishable with the successful probability one. The optimal parameters are $\boldsymbol{\phi} = (0,-0.31973,0,0,-1.5708,0)$ and $\boldsymbol{\theta} = (0,-1.3065,0,0)$. We note that $\frac{\pi}{2} \simeq 1.5708$, and we can find that the distance is one for any values of $\phi_2$ and $\theta_2$. 

The two ground states $\ket{\Phi_{\rm ghz}}$ and $\ket{\Phi_+}$ are not orthogonal. Therefore, they are not fully distinguishable. The trace distance between the two quantum states is 
\begin{eqnarray}
D(\ket{\Phi_{\rm ghz}},\ket{\Phi_+}) &=& \sqrt{1-\abs{\braket{\Phi_+}{\Phi_{\rm ghz}}}^2} \notag \\
&=& \sqrt{1-\frac{1}{2^{M-1}}},
\end{eqnarray}
where $M$ is the number of qubits in the ground states. For any measurement setup, the distance between measurement-outcome distributions of two quantum states is never larger than $D(\ket{\Phi_{\rm ghz}},\ket{\Phi_+})$. Therefore, $D(p_{\boldsymbol{\phi},\boldsymbol{\theta}}(\bullet\vert\Phi_{\rm ghz}),p_{\boldsymbol{\phi},\boldsymbol{\theta}}(\bullet\vert\Phi_+)) \leq D(\ket{\Phi_{\rm ghz}},\ket{\Phi_+})$. 

If two ground states are directly measured in the $Z$ basis, the measurement-outcome distributions are $q(\boldsymbol{\mu}\vert\Phi_{\rm ghz}) = \frac{\delta_{\boldsymbol{\mu},\boldsymbol{0}}+\delta_{\boldsymbol{\mu},\boldsymbol{1}}}{2}$ and $q(\boldsymbol{\mu}\vert\Phi_+) = \frac{1}{2^M}$, where $\boldsymbol{0} = (0,0,\ldots,0)$ and $\boldsymbol{1} = (1,1,\ldots,1)$. The distance between the two distributions is 
\begin{eqnarray}
D(q(\bullet\vert\Phi_{\rm ghz}),q(\bullet\vert\Phi_+)) = 1-\frac{1}{2^{M-1}}.
\end{eqnarray}

To distinguish two ground states of five qubits, we take $M=N=5$, i.e.~each layer has five qubits or classical bits. We find optimal parameters $\boldsymbol{\phi}$ and $\boldsymbol{\theta}$ by maximising the distance function 
\begin{eqnarray}
\overline{D}(\boldsymbol{\phi},\boldsymbol{\theta}) = D(p_{\boldsymbol{\phi},\boldsymbol{\theta}}(\bullet\vert\Phi_{\rm ghz}),p_{\boldsymbol{\phi},\boldsymbol{\theta}}(\bullet\vert\Phi_+)).
\label{eq:dis}
\end{eqnarray}
The result is plotted in Fig.~\ref{fig:find}(b), and $\overline{D}$ reaches the quantum upper bound $D(\ket{\Phi_{\rm ghz}},\ket{\Phi_+}) \simeq 0.96824$ at the end of the optimisation. 

If we turn off parameters $\boldsymbol{\phi}$ by setting $\phi_i = 0$ for all $i = 1,2,\ldots,(M+1)N$, we find the optimal $\boldsymbol{\theta}$ by maximising the distance function 
\begin{eqnarray}
\overline{D}(\boldsymbol{\theta}) = D(p_{\boldsymbol{0},\boldsymbol{\theta}}(\bullet\vert\Phi_{\rm ghz}),p_{\boldsymbol{0},\boldsymbol{\theta}}(\bullet\vert\Phi_+)).
\end{eqnarray}
Here, $\boldsymbol{0}$ is the $(M+1)N$-dimensional zero vector. The result is plotted in Fig.~\ref{fig:find}(c), and $\overline{D}$ reaches $0.94792$ at the end of the optimisation, which is lower than $D(\ket{\Phi_{\rm ghz}},\ket{\Phi_+})$ but above $D(q(\bullet\vert\Phi_{\rm ghz}),q(\bullet\vert\Phi_+)) = 0.9375$. 

For the experiment, we use a three-qubit neural network shown in Fig.~5(a), i.e.~$M=2$ and $N=1$, to distinguish two-qubit ground states. We find optimal parameters $\boldsymbol{\phi}$ and $\boldsymbol{\theta}$ by maximising the distance function $\overline{D}(\boldsymbol{\phi},\boldsymbol{\theta})$ [Eq.~(\ref{eq:dis})], and the result is plotted in Fig.~\ref{fig:find}(d). The distance $\overline{D}$ reaches the quantum upper bound $D(\ket{\Phi_{\rm ghz}},\ket{\Phi_+}) \simeq 0.70711$ at the end of the optimisation. The optimal parameters are $\boldsymbol{\phi} = (1.5708,1.5708,-0.78540)$ and $\boldsymbol{\theta} = (0,0)$. We note that $\frac{\pi}{2} \simeq 1.5708$ and $\frac{\pi}{4} \simeq 0.78540$. These parameters are used in the experiment. 

In the experiment of three-qubit neural network implemented on {\it ibmq{\textunderscore}vigo}, we optimise the implementation, i.e.~minimise the number of two-qubit gates, as follows. We can find that only memristive gates $M_\theta$ with $\theta = 0$ are used according to optimal parameters. Each gate $M_0$ can be realised using two controlled-NOT gates, as shown in Fig.~\ref{fig:circuits}(b), which is equivalent to a controlled-NOT gate followed by a SWAP gate. Therefore, we can implement the neural network with optimal parameters as shown in Fig.~\ref{fig:circuits}(f): At the beginning, qubit-0 represents the resistance qubit (i.e.~hidden-layer qubit), qubit-1 and qubit-2 represent current qubits (i.e.~input qubits); To perform the first memristive gate, instead of physically performing the SWAP gate, the roles of qubit-0 and qubit-1 are exchanged after the first controlled-NOT gate, i.e.~now qubit-1 represents the resistance qubit, and qubit-0 represents a current qubit; It is similar for the second memristive gate. The distributions of measurement outcomes obtained in the experiment are shown in Fig.~5(b). The distance between distributions of two ground states is $0.65673$, which is lower than the theoretical value $D(\ket{\Phi_{\rm ghz}},\ket{\Phi_+}) \simeq 0.70711$ but above $D(q(\bullet\vert\Phi_{\rm ghz}),q(\bullet\vert\Phi_+)) = 0.5$.

\end{document}